\documentclass[12pt,letterpaper,floats]{article}
\usepackage{epsfig,color}
\newcommand{\nc}{\newcommand}
\newcommand{\rnc}{\renewcommand}

\makeatletter
\rnc{\theequation}{\thesection.\arabic{equation}}
\@addtoreset{equation}{section}
\makeatother
\nc{\fig}[5]{
\begin{figure}[!htbp]
    \begin{center}
    \leavevmode
    \centerline{
        \includegraphics[width=#1, height=#2]{#3}
        }
    \caption[]{#4}
    \label{#5}
    \end{center}
\end{figure}}
\begin{document}
\begin{flushright}
{\tt hep-th/0504172}
\end{flushright}
\vspace{2mm}
\begin{center}
{{{\Large \textbf{Ghost Condensation in the Brane-World}}}}\\[13mm]
{Robert B. Mann\footnote{%
mann@avatar.uwaterloo.ca} and John J. Oh\footnote{%
j4oh@sciborg.uwaterloo.ca}}\\[3mm]
\textit{Department of Physics, University of Waterloo,\\[0pt]
Waterloo, Ontario, N2L 3G1, Canada}\\[0pt]
\end{center}
\vspace{2mm}
\begin{abstract}
Motivated by the ghost condensate model, we study the
Randall-Sundrum (RS) brane-world with an arbitrary function of the
higher derivative kinetic terms, $\mathcal{P}(X)$, where
$X=-(\nabla \phi )^{2}$. The five-dimensional Einstein equations
reduce to two equations of motion with a constraint between
$\mathcal{P}(X)$ and the five-dimensional cosmological constant on
the brane. For a static extra dimension, $\mathcal{P}(X)$ has
solutions for both a negative kinetic scalar (so called
{\textit{ghost}}) as well as an ordinary scalar field. However
ghost condensation cannot take place. We show that
small perturbations along the extra dimensional radius (the radion) can give rise to
ghost condensation. This produces a
radiation-dominated universe and the vanishing cosmological
constant at late times but destabilizes the radion. This
instability can be resolved by an inclusion of  bulk matter along
$y$-direction, which finally presents a possible explanation of the
late-time cosmic acceleration.
\end{abstract}

\vspace{5mm}


\hspace{10.5cm}{Typeset Using \LaTeX}

\newpage

\setcounter{equation}{0}

\section{Introduction}

\label{sec:I} Physics in spacetimes with extra dimensions yields
many theoretical insights and poses new experimental challenges.
Though the idea of introducing higher dimensions has been around
for more than 80 years (since the Kaluza-Klein proposal), it has
been successfully reinvented since the advent of string theory. In
large part this is due to the general expectation that the
existence of additional spatial coordinates might resolve a number
of problematic issues such as the smallness of cosmological
constant, the hierarchy between energy scales, and the
accelerating expansion of the universe, that consideration becomes
of great interest and importance in the last few years.

>From this perspective the brane-world scenario offers tantalizing new
prospects for addressing puzzling issues rooted in both cosmology and
particle physics. Pioneering work by Randall and Sundrum (RS) \cite{rs}
posited two models with non-flat extra dimensions in which the universe is
regarded as a three-dimensional brane located at a fixed point of an $%
\mathbf{S}^{1}/Z_{2}$ orbifold in five dimensions. The zero modes of the
gravitational field, which turn out to be massless on the branes, can be
trapped on the brane for perturbations around a flat brane geometry. In the
context of string theory, the context of the model arises from the $%
E_{8}\times E_{8}$ heterotic string theory related to an eleven-dimensional
supergravity theory on orbifold $\mathbf{R}^{10} \times \mathbf{\ S}%
^{1}/Z_{2}$ \cite{hw}. The RS brane-world scenario has drawn much attention
because it offers new possibilities for addressing both the gauge hierarchy
problem and the cosmological constant problem. Its rather different
cosmological perspective generated several attempts to recover conventional
cosmology from it. These include radion stabilization mechanisms \cite%
{bdl,cgkt,cgrt,cgs,hbkim,bdel,klmk} and related works associated with
inflation \cite{inflation} and quintessential brane models \cite{diaz}.

On the other hand, Dvali \textit{et al.} suggested a brane-induced model
with a flat large extra dimension in five dimensions \cite{dgps}, which
shows that the theory on a $3$-brane in five dimensional Minkowski
spacetimes gives rise to the four-dimensional Newtonian potential at short
distances while the potential at large distances is that of the
five-dimensional theory. Therefore, the model was regarded as an interesting
attempt at modifying gravity in the infrared (IR) region. In the context of
IR modification of general relativity, another intriguing suggestion is the
\textit{ghost condensation} mechanism \cite{aclm} in which a condensing
ghost field forms a sort of fluid with an equation-of-state parameter, $%
\omega=-1$, where $\omega=p/\rho$, that fills the universe but has different
properties from that of a cosmological constant. The ghost condensate breaks
time-translation symmetry (a kind of Lorentz symmetry breaking), making the
graviton massive in the IR region and giving rise to a stable vacuum state
in spite of a wrong-signed kinetic term. This is somewhat like a
gravitational analogue of the Higgs mechanism, and the model has been
proposed as a candidate that could account for a consistent IR modification
of general relativity and for a connection between inflation and the dark
energy/matter \cite{gc,av,kn}.

Although incorporating ghosts in a cosmological model with extra dimensions
has been suggested before \cite{kkop,pospelov}, a model with condensing ghosts
offers a mechanism for producing matter or vacuum energy from exotic objects
that has not yet been considered. Motivated by the preceding considerations,
in this paper we explore the consequences of combining the brane-world
scenario with the ghost condensate model. We find that ghost condensation is
not possible with a static extra dimension. A dynamic radion field can provide a
mechanism for condensation, but is generically unstable. Another way to
achieve ghost condensation is by introducing bulk matter. This can make the
radion static, but has the effect of reducing the system to the four-dimensional
case studied in ref.\cite{kn}.

In Sec.~\ref{sec:II}, we present the five-dimensional Einstein-Hilbert
action with the cosmological constant, brane tensions, and an arbitrary
function of the scalar kinetic terms, $\mathcal{P}(X)$. The field equation
for $\mathcal{P}(X)$ is investigated. In Sec.~\ref{sec:III}, the brane
junction conditions are imposed and the five-dimensional equations of motion
with a constraint are derived for a static extra dimension. The Friedmann
and the acceleration equations on the brane are derived by following the Bin$%
\acute{\mathrm{e}}$truy-Deffayet-Langlois (BDL) type approach \cite%
{bdl,hbkim,bdel,klmk}. We show that the conservation equations on
the brane are satisfied.
In Sec.~\ref{sec:IV}, we show that small
perturbations of the brane along the extra dimension can give rise
to ghost condensation and the equations at late times can
determine the vacuum as Minkowski spacetime. Although the model
describes the radiation-dominated universe at late times, it is accompanied
by an unstable radion field. We address this issue in
Sec.~\ref{sec:V}, where we show that the instability of the radion
field can be resolved by introducing bulk matter along the
$y$-direction. This bulk matter can make the radion stable,
and reduces the system to that of the four-dimensional case, which has
an inflationary solution for the scale factor and a late-time
cosmic speed-up. This also
offers consistent solutions with the stable radius of extra dimension,
describing a de-Sitter (dS) phase in which the resulting evolution of
the ghost condensate behaves like a cosmological constant with $\omega =
-1$ at late times. Finally, some discussion and comments on our
results are given in Sec.~\ref{sec:VI}.

\section{Setup and Equations of Motion}

\label{sec:II}

Let us consider a five-dimensional spacetime with a $\mathbf{S}^{1}/Z_{2}$
orbifold structure along the fifth direction. The five-dimensional
Einstein-Hilbert action with the matter lagrangian is given by
\begin{equation}
S=\int d^{4}xdy\sqrt{-\hat{g}}\left[ \frac{M_{(5)}^{4}}{2}\hat{R}+\hat{%
\mathcal{L}}_{\mathrm{M}}\right] ,  \label{eq:action}
\end{equation}%
where hatted quantities are five-dimensional and $M_{(5)}^{4}=1/8\pi G_{(5)}$%
, with $G_{(5)}$ the five-dimensional gravitational constant. The most
general cosmological metric ansatz consistent with the orbifold structure is
given by the following line element
\begin{equation}
(ds)^{2}=-n^{2}(t,y)dt^{2}+a^{2}(t,y)h_{ij}dx^{i}dx^{j}+b^{2}(t,y)dy^{2},
\label{eq:metric}
\end{equation}%
where $h_{ij}$ is a $3$-dimensional homogeneous and isotropic induced metric
defined by
\begin{equation}
h_{ij}=\delta _{ij}+\frac{\alpha x_{i}x_{j}}{1-\alpha x^{m}x^{n}\delta _{mn}}%
,  \label{eq:hiimetric}
\end{equation}%
where $\alpha $ represents its spatial curvature which has the value of $%
0,\pm 1$.

>From the metric (\ref{eq:metric}), we obtain
\begin{eqnarray}
\hat{G}_{00} &=&3\left[ \left( \frac{\dot{a}}{a}\right) ^{2}+\frac{\dot{a}%
\dot{b}}{ab}-\frac{n^{2}}{b^{2}}\left( \frac{a^{\prime \prime }}{a}-\frac{%
a^{\prime }b^{\prime }}{ab}+\left( \frac{a^{\prime }}{a}\right) ^{2}\right) +%
\frac{n^{2}}{a^{2}}\alpha \right] ,  \label{eq:einstensor} \\
\hat{G}_{ij} &=&-h_{ij}\frac{a^{2}}{n^{2}}\left[ \left( \frac{\dot{a}}{a}%
\right) ^{2}+2\frac{\ddot{a}}{a}-2\frac{\dot{n}\dot{a}}{na}+2\frac{\dot{a}%
\dot{b}}{ab}-\frac{\dot{n}\dot{b}}{nb}+\frac{\ddot{b}}{b}\right]  \nonumber
\\
&+&h_{ij}\frac{a^{2}}{b^{2}}\left[ \left( \frac{a^{\prime }}{a}\right) ^{2}+2%
\frac{a^{\prime \prime }}{a}-2\frac{a^{\prime }b^{\prime }}{ab}+2\frac{%
n^{\prime }a^{\prime }}{na}-\frac{n^{\prime }b^{\prime }}{nb}+\frac{%
n^{\prime \prime }}{n}\right] -\alpha h_{ij}, \\
\hat{G}_{44} &=&3\left[ \left( \frac{a^{\prime }}{a}\right) ^{2}+\frac{%
a^{\prime }n^{\prime }}{an}-\frac{b^{2}}{a^{2}}\alpha -\frac{b^{2}}{n^{2}}%
\left( \left( \frac{\dot{a}}{a}\right) ^{2}+\frac{\ddot{a}}{a}-\frac{\dot{n}%
\dot{a}}{na}\right) \right] , \\
\hat{G}_{04} &=&3\left[ \frac{n^{\prime }\dot{a}}{na}-\frac{\dot{a^{\prime }}%
}{a}+\frac{a^{\prime }\dot{b}}{ab}\right] ,
\end{eqnarray}%
for the Einstein tensor, where the dot and the prime respectively denote $t$
and $y$ derivatives.

The matter lagrangian $\hat{\mathcal{L}}_{\mathrm{M}}$ includes the brane
tension, $\hat{\mathcal{L}}_{\mathrm{M}}^{\mathrm{brane}}$, the
five-dimensional cosmological constant, $\hat{\mathcal{L}}_{\mathrm{M}%
}^{\Lambda _{(5)}}$, the bulk matter along the $y$-direction,
$\hat{\mathcal{L}}_{\mathrm{M}}^{\mathrm{bulk}}$, and an arbitrary function of the kinetic terms, $\hat{%
\mathcal{L}}_{\mathrm{M}}^{\mathrm{\mathcal{P}}}$,
\begin{equation}
\hat{\mathcal{L}}_{\mathrm{M}}=\hat{\mathcal{L}}_{\mathrm{M}}^{\mathrm{brane}%
}+\hat{\mathcal{L}}_{\mathrm{M}}^{\Lambda _{(5)}}+\hat{\mathcal{L}}_{\mathrm{M}}^{\mathrm{bulk}}+\hat{\mathcal{L}}_{\mathrm{%
M}}^{\mathrm{\mathcal{P}}}  \label{eq:matter}
\end{equation}%
where the lagrangians of the brane tension and cosmological constant yield
in turn simple expressions for their energy-momentum
\begin{eqnarray}
\hat{T}_{~~N}^{M~\Lambda _{(5)}}&+&\hat{T}_{~~N}^{M~\mathrm{bulk}}+\hat{T}_{~~N}^{M~\mathrm{brane}}=diag(-\rho
_{\Lambda },p_{\Lambda },p_{\Lambda },p_{\Lambda },p_{\Lambda
}) \nonumber \\&+& M_{(5)}^4 diag(0,0,0,0,p_{5})+\sum_{u=0,\pi}\frac{\delta (y_{u})}{b}diag(-\rho _{u},p_{u},p_{u},p_{u},0)
\label{eq:emt_brane}
\end{eqnarray}%
where $u$ represents the visible ($u=0$) and hidden ($u=\pi$) branes~\cite%
{rs}, and we have $\rho _{\Lambda }=-p_{\Lambda }=\Lambda _{(5)} $, where $%
\Lambda _{(5)}<0$. In addition, $p_{5}$ is bulk matter along the
$y$-direction that is responsible for the radion stabilization \cite{hbkim,kkop}.
Note that the indices $M,N,\cdots $ and $\mu ,\nu ,\cdots
$ run from $0$ to $4$ and $0$ to $3$, respectively. And we denote $%
i,j,\cdots =1,2,3$, which represents the spatial coordinates on the
three-dimensional brane.

On the other hand $\hat{\mathcal{L}}_{\mathrm{M}}^{\mathrm{\mathcal{P}}}$,
the lagrangian for $\mathcal{P}(X)$, can be written as
\begin{equation}
\hat{\mathcal{L}}_{\mathrm{M}}^{\mathrm{\mathcal{P}}}=M_{(5)}^{4}\mathcal{P}%
(X),  \label{eq:ghostmatter}
\end{equation}%
where $\mathcal{P}(X)$ is an arbitrary function of $X=-\partial _{M}\phi
\partial ^{M}\phi $ that describes the higher derivative terms of the
kinetic scalar field. This type of matter was considered in connection with {%
k-essence} \cite{kessence} and {ghost condensation} \cite{aclm}. The
energy-momentum tensor of eq. (\ref{eq:ghostmatter}) is
\begin{equation}
\hat{T}_{MN}^{\mathrm{\mathcal{P}}}=M_{(5)}^{4}\left( \mathcal{P}(X)\hat{g}%
_{MN}+2\frac{\partial \mathcal{P}(X)}{\partial X}\partial _{M}\phi \partial
_{N}\phi \right).  \label{eq:emtenghost}
\end{equation}%
Einstein's equation is easily obtained from eqs. (\ref{eq:action}), (\ref%
{eq:emt_brane}), and (\ref{eq:emtenghost}),
\begin{equation}
\hat{G}_{MN}=\frac{1}{M_{(5)}^{4}}{\hat{T}}_{MN}=\frac{1}{M_{(5)}^{4}}({\hat{%
T}}_{MN}^{\Lambda _{(5)}}+{\hat{T}}_{MN}^{\mathrm{bulk}}+{\hat{T}}_{MN}^{\mathrm{brane}}+{\hat{T}}_{MN}^{%
\mathrm{\mathcal{P}}}).  \nonumber  \label{eq:einsteq}
\end{equation}

The field $\mathcal{P}(X)$ satisfies
\begin{equation}
\partial _{M}\left( \sqrt{-\hat{g}}\frac{\partial \mathcal{P}(X)}{\partial X}%
\partial ^{M}\phi \right) =0,  \label{eq:eqnmotghost}
\end{equation}%
which has the solution
\begin{equation}
\frac{a^{3}b}{n}\partial _{X}\mathcal{P}(\dot{\phi}^{2})\dot{\phi}={const},
\label{eq:soleqmot}
\end{equation}%
for $\phi =\phi (t)$ when $\alpha =0$. Writing $a^{2}(t,y)=n^{2}(t,y){a_{0}}%
^{2}(t)$ and $n^{2}(t,y)=e^{-2b(t,y)\sigma (y)}$ (reflecting the \textit{%
warped geometry}), the metric (\ref{eq:metric}) becomes
\begin{equation}
(ds)^{2}=e^{-2b\sigma (y)}\left( -dt^{2}+{a_{0}}^{2}(t)h_{ij}dx^{i}dx^{j}%
\right) +b^{2}(t,y)dy^{2},  \label{eq:redmetric}
\end{equation}%
and eq. (\ref{eq:soleqmot}) can be written as
\begin{equation}
be^{-2b\sigma }\partial _{X}\mathcal{P}(\dot{\phi}^{2})\dot{\phi}=\frac{const%
}{{a_{0}}^{3}}.  \label{eq:soleqnmot3}
\end{equation}%
Since we expect that the scale factor $a_{0}$ on the brane goes to infinity
at late times, at least one of the following three scenarios
\begin{equation}
\dot{\phi}\rightarrow 0~~\mathrm{or}~~\partial _{X}\mathcal{P}(\dot{\phi}%
^{2})\rightarrow 0~~\mathrm{or}~~be^{-2b}\rightarrow 0  \label{eq:3sols}
\end{equation}%
must ensue. The first option is the case that the scalar velocity goes to
zero whereas the second leads to ghost condensation at late times; both of
these options have been previously considered \cite{aclm,kn}.

The third option arises from the five-dimensional brane-world hypothesis.
This case implies that the time-evolution of the extra dimension, $b(t,y)$,
has solutions $b=0$ or $b=\infty$ at late times. Either of these solutions
automatically satisfies the field equation at late times. To address the
gauge hierarchy problem, the RS1 model assumes a very small extra dimension,
congruent with $b\rightarrow 0$, while the $b\rightarrow \infty$ limit is
congruent with the RS2 model, which has a semi-infinite extra dimension \cite%
{rs}.

In the brane-world scenario, the position of a brane can be stabilized by
the radion stabilization mechanism \cite{cgkt}. Once the radius of the extra
dimension is fixed, it has a static configuration, $\dot{b}=0$. We shall
explore this option in the next section.

\section{Ghosts and a Static Extra Dimension}

\label{sec:III}

\subsection{Brane Junction Condition and the Ghost}

\label{sub:3.1}
In this section, we consider a model with a static extra dimension and $p_{5}=0$.
Since there exist nontrivial topological defects such as 3-branes orthogonal
to the fifth direction, an appropriate junction condition is required to
resolve the discontinuity. These conditions can be imposed by integrating $%
(00)$- and $(11)$-components of the Einstein's equation, yielding
\begin{eqnarray}
&& [a^{\prime }]=-\frac{a_{u}b_{u}}{3M_{(5)}^{4}}\rho _{u},
\label{eq:junctiona} \\
&& [n^{\prime }]=\frac{n_{u}b_{u}}{3M_{(5)}^{4}}(2\rho _{u}+3p_{u})
\label{eq:junctionn}
\end{eqnarray}
on the branes at $y=0$ and $y=\pi $, where the prime denotes a derivative
with respect to $y$ and $[a^{\prime}]=a^{\prime}(y_{+u}) -
a^{\prime}(y_{-u}) $. Using the junction condition on the branes at $y=0$,
the static extra dimension ($\dot{b}=0$), and the gauge fixing of $b=r_{c} =
const.$ and $n_{0}\equiv n(t,0)=1$, the evolution equations on the brane
become
\begin{eqnarray}
&&H_{0}^{2}+\frac{\alpha}{a_{0}^{2}} =\frac{1}{3}\left( 2X\partial _{X}%
\mathcal{P}(X)-\mathcal{P}(X)\right) +\frac{\Lambda_{(5)}}{3M_{(5)}^{4}} +
\frac{\rho_{0}^2}{36M_{(5)}^8},  \label{eq:friedmann} \\
&&H_{0}^{2}+\frac{\alpha}{a_{0}^{2}} +2\frac{\ddot{a}_{0}}{a_{0}} =-\mathcal{%
P}(X)+\frac{\Lambda_{(5)}}{M_{(5)}^{4}} + \frac{\rho_{0}^2}{36M_{(5)}^8} -
\frac{2\rho_{0}}{36M_{(5)}^8}(2\rho_{0}+3p_{0}),  \label{eq:evolution} \\
&&H_{0}^{2}+\frac{\alpha}{a_{0}^{2}} +\frac{\ddot{a}_{0}}{a_{0}} =-\frac{1}{3%
}\mathcal{P}(X)+\frac{\Lambda_{(5)}}{3M_{(5)}^{4}} + \frac{\rho_{0}^2}{%
36M_{(5)}^8} - \frac{\rho_{0}}{36M_{(5)}^8}(2\rho_{0}+3p_{0}) ,
\label{eq:extra}
\end{eqnarray}%
where $H_{0}^{2}=\dot{a}_{0}^{2}/a_{0}^{2}$ is the Hubble's parameter
defined on the brane at $y=0$. Combining the above three equations yields a
simple relation between the function $\mathcal{P}(X)$ and the cosmological
constant,
\begin{equation}
\mathcal{P}(X)-X\partial _{X}\mathcal{P}(X)=\frac{\Lambda_{(5)}}{M_{(5)}^{4}}%
,  \label{eq:gcrelation}
\end{equation}%
which has the general solution
\begin{equation}
\mathcal{P}(X)=\frac{\Lambda _{(5)}}{M_{(5)}^{4}}+CX,  \label{eq:gensolution}
\end{equation}%
where $C$ is an integration constant. Note that if $C$ is positive, this
describes an ordinary scalar field (stiff matter) while negative $C$
represents a ghost scalar field. Since there is no restriction on the choice
of sign for $C$, these two solutions are both allowed. Equation (\ref%
{eq:gensolution}) reduces the Friedmann and acceleration equations to%
\begin{eqnarray}
&&H_{0}^{2}+\frac{\alpha}{a_{0}^{2}} = \frac{1}{3}{CX} + \frac{\rho_{0}^2}{%
36M_{(5)}^8},  \label{eq:reducedeqn} \\
&&\frac{\ddot{a}_{0}}{a_{0}}= -\frac{2}{3}{CX} - \frac{\rho_{0}}{36M_{(5)}^8}%
(2\rho_{0}+3p_{0}).  \label{eq:reducedeqn2}
\end{eqnarray}

We see from (\ref{eq:gensolution}) that the brane-world does not allow
higher derivative kinetic terms if the extra dimension is static. We shall
see in sec. \ref{sec:IV} that removing this constraint (i.e. assuming a
non-static extra dimension) gives rise to ghost condensation.

\subsection{Ghosts on the Brane}

\label{sub:3.2}

In this section, we shall derive the Friedmann and the acceleration
equations on the brane at $y=0$ by imposing the brane junction condition
using the BDL approach \cite{bdl,hbkim,bdel,klmk}. To investigate the
Friedmann equation on the brane at $y=0$, we define
\begin{equation}
J(t,y)\equiv \frac{(a^{\prime }a)^{2}}{b^{2}}-\frac{(\dot{a}a)^{2}}{n^{2}}%
-\alpha a^{2}.  \label{eq:jtomega}
\end{equation}%
The $(00)$-and $(44)$-components of the Einstein's equation in the bulk are
then
\begin{eqnarray}
& &J^{\prime}(t,y) = \frac{1}{6} \partial_{y}(a^4)\frac{\hat{T}^{t}_{~t}(t,y)%
}{M_{(5)}^4},  \label{eq:redineq1} \\
& &\dot{J}(t,y) = \frac{1}{6}\partial_{t}(a^4)\frac{\hat{T}^{y}_{~y}(t,y)}{%
M_{(5)}^4},  \label{eq:redineq2}
\end{eqnarray}
where $\hat{T}^{M}_{~N} = \hat{T}^{M~\Lambda_{(5)}}_{~N} + \hat{T}^{M~{%
\mathcal{P}}}_{~N}$ in the bulk. It is easy to show that above equations
satisfy the continuity condition, $\partial_{t}\partial_{y}J(t,y)=%
\partial_{y}\partial_{t}J(t,y)$, by using the $(04)$-component of Einstein's
equation and the conservation of the energy-momentum tensor (or field
equation for ${\mathcal{P}}(X)$). Notice that eqs. (\ref{eq:redineq1}) and (%
\ref{eq:redineq2}) can be simply solved as
\begin{eqnarray}
J(t,y) &=& \frac{a^4}{6M_{(5)}^4}\hat{T}^{t}_{~t} - \frac{1}{6M_{(5)}^4}
\int dy a^4 \partial_{y} \hat{T}^{t~{\mathcal{P}}}_{~t} + \Omega(t)
\label{eq:solved} \\
&=& \frac{a^4}{6M_{(5)}^4}\hat{T}^{y}_{~y} - \frac{1}{6M_{(5)}^4} \int dt
a^4 \partial_{t} \hat{T}^{y~{\mathcal{P}}}_{~y} + \tilde{\Omega}(y),
\label{eq:solved2}
\end{eqnarray}
where $\Omega(t)$ and $\tilde{\Omega}(y)$ are integration functions that
respect the relation
\begin{equation}  \label{eq:integratconst}
\Omega(t) = \frac{a^4}{6M_{(5)}^4} (\hat{T}^{y~{\mathcal{P}}}_{~y}-\hat{T}%
^{t~{\mathcal{P}}}_{~t}) + \frac{1}{6M_{(5)}^4}\left(\int dy a^4 \partial_{y}%
\hat{T}^{t~{\mathcal{P}}}_{~t} - \int dt a^4\partial_{t}\hat{T}^{y~{\mathcal{%
P}}}_{~y}\right) + \tilde{\Omega}(y).
\end{equation}

On the other hand, the $(04)$-component of the Einstein's equation is
\begin{equation}
\frac{n^{\prime }\dot{a}}{na}-\frac{\dot{a}^{\prime }}{a}=0,
\label{eq:04com}
\end{equation}%
for a static radius of extra dimension, $\dot{b}=0$, which implies that $%
\dot{a}(t,y)=\nu (t)n(t,y)$.

>From eq. (\ref{eq:solved}), we have
\begin{equation}  \label{eq:7}
\frac{a^{\prime 2}}{b^2}= \nu^2(t) + \alpha - k^2a^2 + \frac{a^2}{6M_{(5)}^4}%
\hat{T}^{t~{\mathcal{P}}}_{~t} - \frac{1}{6M_{(5)}^4 a^2} \int dy a^4
\partial_{y} \hat{T}^{t~{\mathcal{P}}}_{~t} + \frac{\Omega(t)}{a^2},
\end{equation}
where $k^2 \equiv \Lambda_{(5)}/6M_{(5)}^4$. If we define $\hat{\mathcal{G}}%
^{\mathcal{P}}(t,y) \equiv {\hat{T}^{t~{\mathcal{P}}}_{~t}}/{6M_{(5)}^4}$,
then eq. (\ref{eq:7}) becomes
\begin{equation}  \label{eq:10}
\int bk dy = \int \frac{da}{\sqrt{\frac{\nu^2+\alpha}{k^2} - a^2 + \frac{a^2%
\hat{\mathcal{G}}^{\mathcal{P}}}{k^2} + \frac{\Omega}{k^2a^2}}}\left(1-
\frac{\frac{1}{a^4k^2} \int dy a^4\partial_{y}\hat{\mathcal{G}}^{\mathcal{P}}%
}{\frac{\nu^2+\alpha}{k^2} - a^2 + \frac{a^2\hat{\mathcal{G}}^{\mathcal{P}}}{%
k^2} + \frac{\Omega}{k^2a^2}}\right)^{-1/2}.
\end{equation}
Note that the RHS of eq. (\ref{eq:10}) is the integration with respect to
only $a$. However, because of the integration term in the square-root, we
cannot get an exact solution of eq. (\ref{eq:10}) even if ${\hat{\mathcal{G}}%
^{\mathcal{P}}}$ does not depend upon $a$. However, at this stage, if we
take a near-brane limit $y\rightarrow 0$, then $\partial_{y} \hat{\mathcal{G}%
}^{\mathcal{P}}$ vanishes since $n(t,y)|_{y\rightarrow 0} \rightarrow
n(t,0)=n_{0}(t) = 1$ by gauge fixing. Therefore, we can approximately use
the usual BDL approach in this sense.

Integrating eq. (\ref{eq:10}) leads to the following solution for the scale
factor
\begin{equation}
a^{2}(t,y)=\frac{W(t)e^{2\sqrt{{\mathcal{F}}}bky}}{2\sqrt{{\mathcal{F}}}}-%
\frac{\nu ^{2}(t)+\alpha }{2{\mathcal{F}}k^{2}}+\frac{e^{-2\sqrt{{\mathcal{F}%
}}bky}}{2\sqrt{{\mathcal{F}}}k^{2}W(t)}\left[ \left( \frac{\nu
^{2}(t)+\alpha }{2\sqrt{{\mathcal{F}}}k}\right) ^{2}-\Omega (t)\right] ,
\label{eq:solscale}
\end{equation}%
where ${\mathcal{F}} = (\hat{\mathcal{G}}^{\mathcal{P}} - k^2)/k^2$ and $W(t)
$ is an arbitrary function that arises from the integration. Here we define
the functions on the brane at $y=0$ as $a_{0}(t)\equiv a(t,0)$ and $\dot{a}%
_{0}(t)\equiv \dot{a}(t,0)=\nu \left( t\right) $, and fix the gauge $n(t,0)=1
$. Then, eq. (\ref{eq:solscale}) on the brane at $y=0$ can be written by
\begin{equation}
W(t)=\sqrt{{\mathcal{F}}}a_{0}^{2}\left[ 1+\frac{\dot{a}_{0}^{2}+\alpha }{2{%
\mathcal{F}}k^{2}a_{0}^{2}}\pm \left( 1+\frac{1}{{\mathcal{F}}k^{2}}\left(
\frac{\dot{a}_{0}^{2}+\alpha }{a_{0}^{2}}+\frac{\Omega (t)}{a_{0}^{4}}%
\right) \right) ^{1/2}\right] .  \label{eq:branescale2}
\end{equation}%
We, therefore, obtain the exact solution of the scale factor expressed in
terms of the brane scale factor, $a_{0}$,

\begin{eqnarray}
&&a^{2}(t,y)=a_{0}^{2}\left[ \left( 1+\frac{\dot{a}_{0}^{2}+\alpha }{2{%
\mathcal{F}}k^{2}a_{0}^{2}}\right) \cosh (2\sqrt{{\mathcal{F}}}kby)\right.
\nonumber  \label{eq:finalsol} \\
&\pm &\left. \left( 1+\frac{1}{{\mathcal{F}}k^{2}}\left( \frac{\dot{a}%
_{0}^{2}+\alpha }{a_{0}^{2}}+\frac{\Omega (t)}{a_{0}^{4}}\right) \right)
^{1/2}\sinh (2\sqrt{{\mathcal{F}}}kby)-\frac{\dot{a}_{0}^{2}+\alpha }{2{%
\mathcal{F}}a_{0}^{2}k^{2}}\right] .
\end{eqnarray}%
{}From eqs. (\ref{eq:junctiona}) and (\ref{eq:finalsol}), the Friedmann
equation on the brane is obtained by
\begin{equation}
H_{0}^{2}+\frac{\alpha }{a_{0}^{2}}=k^{2}-\frac{1}{6}({\mathcal{P}}(X) -
2X\partial_{X}{\mathcal{P}}(X)) +\frac{\rho _{0}^{2}}{36M_{(5)}^{8}}-\frac{%
\Omega _{0}}{a_{0}^{4}},  \label{eq:friedmannbranes}
\end{equation}%
where $H_{0}=\dot{a}_{0}/a_{0}$ is a Hubble's parameter defined on the brane
and $\Omega _{0}$ is the value of $\Omega $ at $y=0$. Since we have eq. (\ref%
{eq:gcrelation}) on the brane at $y=0$, eq. (\ref{eq:friedmannbranes}) can
be rewritten by
\begin{equation}
H_{0}^{2}+\frac{\alpha }{a_{0}^{2}}=-k^{2}+\frac{1}{6}{\mathcal{P}}(X)+
\frac{\rho _{0}^{2}}{36M_{(5)}^{8}}-\frac{\Omega _{0}}{a_{0}^{4}}.
\label{eq:friedmannbrane}
\end{equation}%
All terms on the right-hand-side except the second term are familiar terms
from the brane-world cosmological scenario.

The acceleration equation on the brane $y=0$ can be obtained by combining
the ($44$)-component of eq. (\ref{eq:einsteq}) and eqs. (\ref{eq:junctiona}%
), (\ref{eq:junctionn}), (\ref{eq:integratconst}), and (\ref%
{eq:friedmannbrane}),
\begin{equation}
\frac{\ddot{a}_{0}}{a_{0}}=3k^{2}-\frac{1}{2}{\mathcal{P}}(X)-\frac{\rho _{0}%
}{36M_{(5)}^{8}}(2\rho _{0}+3p_{0})+\frac{\Omega _{0}}{a_{0}^{4}}.
\label{eq:accelbrane}
\end{equation}%
In addition, it is easy to show that eqs. (\ref{eq:friedmannbrane}) and (\ref%
{eq:accelbrane}) satisfy the matter conservation equation on the brane,
\begin{equation}
\dot{\rho}_{0}=-3H_{0}(\rho _{0}+p_{0}).  \label{eq:mattercons}
\end{equation}%
Since we have $\mathcal{P}(X)=6k^{2}+CX$ in eq. (\ref{eq:gensolution}), eq. (%
\ref{eq:friedmannbrane}) shows that the usual Friedmann equation can be
described by taking $C$ to be a positive constant (ordinary scalar field).
Eqs. (\ref{eq:friedmannbrane}) and (\ref{eq:accelbrane})
indicate that the contribution of bulk matter to both equations can be
described as a form of generalized bulk matter $\mathcal{P}(X)$, which is
associated with the generalized {\it comoving mass} of the bulk fluid
\cite{at}.

\subsection{Consistent Conservation Equation}

\label{sub:3.3} We can obtain a conservation law for $X$ from eqs. (\ref%
{eq:reducedeqn}) and (\ref{eq:reducedeqn2}) for consistent equations of
motion. First, multiplying both sides of eq. (\ref{eq:reducedeqn}) by $%
a_{0}^{2}$, differentiating them with $t$, dividing by $2\dot{a}_{0}a_{0}$,
we obtain
\begin{equation}
\frac{\ddot{a}_{0}}{a_{0}}=\frac{1}{3}CX+\frac{\rho _{0}^{2}}{36M_{(5)}^{8}}+%
\frac{C\dot{X}}{6H_{0}}+\frac{\rho _{0}\dot{\rho _{0}}}{36M_{(5)}^{8}H_{0}}.
\label{eq:consrvint}
\end{equation}%
This obviously should be equivalent to eq. (\ref{eq:reducedeqn2}), which
yields the relation,
\begin{equation}
C\left( X+\frac{\dot{X}}{6H_{0}}\right) =-\frac{\rho _{0}}{36M_{(5)}^{8}H_{0}%
}\left( \dot{\rho _{0}}+3H_{0}(\rho _{0}+p_{0})\right) .  \label{eq:consvrel}
\end{equation}%
Since the RHS vanishes due to the conservation equation between brane
tension and pressure, (\ref{eq:mattercons}), we have a conservation equation
for $X$
\begin{equation}
\dot{X}=-6H_{0}X.  \label{eq:consvX}
\end{equation}%
It is easy to show Eq. (\ref{eq:consvX}) has the solution $X=X_{0}/a_{0}^{6}$%
, where $X_{0}$ is an integration constant. Alternatively we can simplify
eq. (\ref{eq:integratconst}) when $y=0$, obtaining $\Omega
_{0}=-CX_{0}/6a_{0}^{2}$ (assuming $\tilde{\Omega}(0)=0$). Insertion of this
into eqs. (\ref{eq:friedmannbrane}) and (\ref{eq:accelbrane}) produces
\begin{eqnarray}
&&H_{0}^{2}+\frac{\alpha }{a_{0}^{2}}=\frac{CX_{0}}{3a_{0}^{6}}+\frac{\rho
_{0}^{2}}{36M_{(5)}^{8}},  \label{eq:friedbrane} \\
&&\frac{\ddot{a}_{0}}{a_{0}}=-\frac{2CX_{0}}{3a_{0}^{6}}-\frac{\rho _{0}}{%
36M_{(5)}^{8}}(2\rho _{0}+3p_{0}).  \label{eq:accelbranes}
\end{eqnarray}%
Note that these equations are coincident with eqs. (\ref{eq:reducedeqn}) and
(\ref{eq:reducedeqn2}). Since the $X$ field does not couple to the brane
tension and pressure, the conservation equations between $X$ and $\rho _{0}$
(or/and $p_{0}$) are decoupled as shown before in eqs. (\ref{eq:mattercons})
and (\ref{eq:consvX}).

Assuming $p_{0}=\omega_{b}\rho_{0}$, where $\omega_{b}$ is an
equation-of-state parameter of the 3-brane, eq. (\ref{eq:mattercons}) has
the solution
\begin{equation}  \label{eq:branesol}
\rho_{0} = \bar{\rho}_{0} a_{0}^{-3(1+\omega_{b})},
\end{equation}
where $\bar{\rho}_{0}$ is an integration constant whose value is negative,
since the brane located at $y=0$ is a negative tension brane in the RS1
model \cite{rs} . Therefore, we finally get the Friedmann and the
acceleration equations on the brane in terms of the scale factor $a_{0}(t)$
for the static extra dimension
\begin{eqnarray}
& & H_{0}^2 + \frac{\alpha}{a_{0}^2} = \frac{CX_{0}}{3a_{0}^6} + \frac{\bar{%
\rho}_{0}^2}{36M_{(5)}^8 a_{0}^{6(1+\omega_{b})}},  \label{eq:friedbrane1} \\
& & \frac{\ddot{a}_{0}}{a_{0}} = - \frac{2CX_{0}}{3a_{0}^6} - \frac{\bar{\rho%
}_{0}^2}{36M_{(5)}^8a_{0}^{6(1+\omega_{b})}} (2+3\omega_{b}).
\label{eq:accelbranes1}
\end{eqnarray}

We see from the above that for $C<0$, we obtain acceleration in the brane.
Unfortunately we cannot obtaining a condensing ghost field as eq. (\ref%
{eq:gensolution}) shows. In the next section we address this problem.

\section{Ghost Condensation From Radion Perturbation?}

\label{sec:IV}

The simple ghost and scalar field solutions obtained in section \ref{sec:III}
by assuming a static extra dimension do not lead to ghost condensation in
the brane-world. In this section consider the possibility that a small perturbation of
the radion (i.e. a non-static radius of the extra dimension) can give rise
to ghost condensation. A ghost condensate can be regarded as a new sort of
fluid that fills the universe \cite{aclm}. Its essential property is that it
has a non-vanishing time-dependent vacuum expectation value(VEV). Even
though $\mathcal{P}(X)$ has a wrong-signed kinetic term, there exists a
stable vacuum state with $<\phi >\neq 0$. More precisely, the field equation
of $\mathcal{P}(X)$ can have a solution, $\phi =ct$, where $c$ is a
dimensionless constant, provided that there exists a solution, $\partial
_{t}\phi =\mathrm{constant}$. By a small perturbation about this solution, $%
\phi =ct+\pi $, the lagrangian for quadratic fluctuations as shown in ref. %
\cite{aclm} is given by
\begin{equation}
\mathcal{L}=M^{4}\left[ \left( {\partial _{X}\mathcal{P}(c^{2})}+2c^{2}{%
\partial _{X}^{2}\mathcal{P}(c^{2})}\right) \dot{\pi}^{2}-{\partial _{X}%
\mathcal{P}(c^{2})}(\nabla \pi )^{2}\right] ,  \label{eq:quadfluct}
\end{equation}%
where we have discarded the term $2c\dot{\pi}\partial _{X}P(c^{2})$ since it
is a total derivative. The usual signs can be recovered when $c$ is such
that
\begin{equation}
\partial _{X}\mathcal{P}(c^{2})>0,~~\partial _{X}\mathcal{P}%
(c^{2})+2c^{2}\partial _{X}^{2}\mathcal{P}(c^{2})>0.  \label{eq:ghostrel}
\end{equation}%
The typical shape and region we are considering are shown in Fig.\ref%
{fig:1}.

Let us consider a small\thinspace\ radion perturbation $b=r_{c}+r_{0}(t)$,
where $r_{c}$ is a constant and $r_{0}<<r_{c}$. Using the metric (\ref%
{eq:redmetric}), Einstein's equations on the brane at $y=0$ can rewritten as
\begin{eqnarray}
&&{H}_{0}^{2}+{H}_{0}\frac{\dot{r}_{0}}{r_{c}}+\frac{\alpha }{{a_{0}}^{2}}=%
\frac{1}{3}\left[ 2X\partial _{X}\mathcal{P}(X)-\mathcal{P}(X)\right]
+2k^{2}+\frac{\bar{\rho}_{0}^{2}}{36M_{(5)}^{8}a_{0}^{6(1+\omega _{b})}},
\label{eq:radtt} \\
&&{H}_{0}^{2}+2\left( \frac{\ddot{a}_{0}}{{a}_{0}}+{H}_{0}\frac{\dot{r}_{0}}{%
r_{c}}\right) +\frac{\alpha }{a_{0}^{2}}  \nonumber \\
&=&-\frac{\ddot{r}_{0}}{r_{c}}+6k^{2}-\mathcal{P}(X)+\frac{\bar{\rho}_{0}^{2}%
}{36M_{(5)}^{8}a_{0}^{6(1+\omega _{b})}}-\frac{2\bar{\rho}_{0}^{2}}{%
36M_{(5)}^{8}a_{0}^{6(1+\omega _{b})}}(2+3\omega _{b}),  \label{eq:radii} \\
&&{H}_{0}^{2}+\frac{\ddot{a}_{0}}{{a}_{0}}+\frac{\alpha }{{a}_{0}^{2}}%
=2k^{2}-\frac{1}{3}\mathcal{P}(X)+\frac{\bar{\rho}_{0}^{2}}{%
36M_{(5)}^{8}a_{0}^{6(1+\omega _{b})}}-\frac{\bar{\rho}_{0}^{2}}{%
36M_{(5)}^{8}a_{0}^{6(1+\omega _{b})}}(2+3\omega _{b}),  \label{eq:rad44}
\end{eqnarray}
where we work to leading order in $r_{0}$ and set $\sigma(0) = 0$ (or $%
n_{0}=n(t,0) = 1$) on the brane. Combining these equations yields the
following constraint between $\mathcal{P}(X)$ and the small perturbation of
the radion field,
\begin{equation}
\frac{2}{3}(X\partial _{X}\mathcal{P}(X)-\mathcal{P}(X)+6k^{2})=3H_{0}\frac{%
\dot{r}_{0}}{r_{c}}+\frac{\ddot{r}_{0}}{r_{c}}.  \label{eq:combrel}
\end{equation}

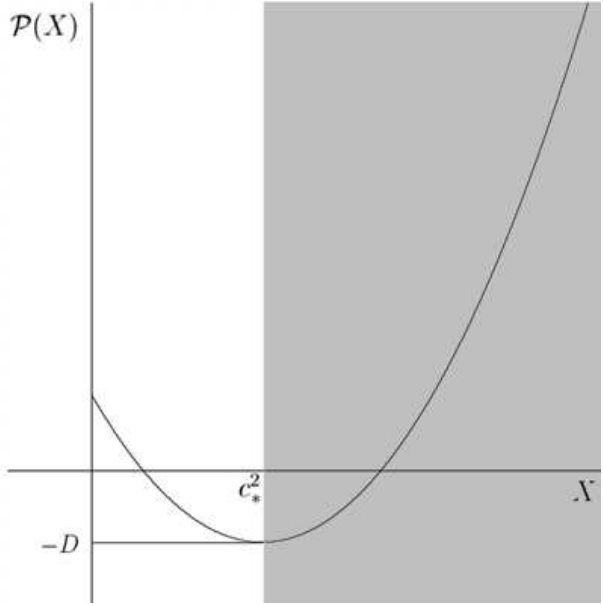
\fig{8cm}{8cm}{fig1}{\small \it {The typical shape of
$\mathcal{P}(X)$. A stability analysis \cite{av} indicates that
the shaded region is the one in which the ghost field behaves like
an ordinary matter field and is stable, whereas the unshaded
region is classically or quantum mechanically
  unstable. }}{fig:1} 

If we assume that the condensing ghost field approaches a stable vacuum at late
times, then $\partial _{X}\mathcal{\ P}(c_{\ast }^{2})\rightarrow 0$
and $\mathcal{P}(c_{\ast }^{2})\rightarrow -D$ as shown in Fig.\ref%
{fig:1}. Defining $6k_{0}^{2}=D+6k^{2}$, Einstein's equations become in
this limit
\begin{eqnarray}
&&{H}_{0}^{2}+\frac{\alpha }{{a_{0}}^{2}}=-{H}_{0}\frac{\dot{r}_{0}}{r_{c}}%
+2k_{0}^{2}+ \frac{\bar{\rho}_{0}^2}{36M_{(5)}^8 a_{0}^{6(1+\omega_{b})}},
\label{eq:reradtt} \\
&&{H}_{0}^{2}+\frac{\alpha }{a_{0}^{2}}  \nonumber \\
&&=-2\left( \frac{\ddot{a}_{0}}{{a}_{0}}+{H}_{0}\frac{\dot{r}_{0}}{r_{c}}%
\right) -\frac{\ddot{r}_{0}}{r_{c}}+6k_{0}^{2}+ \frac{\bar{\rho}_{0}^2}{%
36M_{(5)}^8 a_{0}^{6(1+\omega_{b})}} - \frac{2\bar{\rho}_{0}^2}{%
36M_{(5)}^8a_{0}^{6(1+\omega_{b})}} (2+3\omega_{b}),  \label{eq:reradii} \\
&&{H}_{0}^{2}+\frac{\alpha }{{a}_{0}^{2}}=-\frac{\ddot{a}_{0}}{{a}_{0}}%
+2k_{0}^{2}+ \frac{\bar{\rho}_{0}^2}{36M_{(5)}^8 a_{0}^{6(1+\omega_{b})}} -
\frac{\bar{\rho}_{0}^2}{36M_{(5)}^8a_{0}^{6(1+\omega_{b})}} (2+3\omega_{b}),
\label{eq:rerad44}
\end{eqnarray}%
and combining above three equations leads to two
options: either $\dot{r}_{0}=0$ or $\ddot{a}_{0}/\dot{a}_{0}=0$.

For $\ddot{a}_{0}/\dot{a}_{0}=0$, we find that $a_{0}(t)={\mathcal{A}_{1}}t+{%
\mathcal{A}_{2}}$ where $\mathcal{A}_{1}$ and $\mathcal{A}_{2}$ are
constants, which determines the solution for $r_{0}(t)$ but yields two
options for making this so: $\omega _{b}=-1,-1/3 $, depending upon
the brane tension. These two choices of $\omega _{b}=-1$ and
$\omega_{b}=-1/3$ produce the trivial equation $H_{0}^{2}+{\alpha }/{a_{0}^{2}}=0$.
Hence setting $\ddot{a}_{0}/\dot{a}_{0}=0$ does not yield any
dynamical evolution of the scale factor since the effective energy
density vanishes.

On the other hand, for $\dot{r}_{0}=0$, we obtain $k_{0}^{2}=0$
or $D=-6k^{2}$, which describes a
vanishing cosmological constant at late times. The intriguing point is that
the initially arbitrary value of $D$ is determined by the consistency of
Einstein's equations, leading to a vanishing cosmological constant ($%
k_{0}^{2}=0$) on the brane at late times. In this case, eq. (\ref{eq:reradtt}%
) has solutions of the form for $\alpha=0$,
\begin{eqnarray}
a_{0}(t) &=& \mathcal{C}_{1} e^{\frac{\bar{\rho}_{0}}{6M_{(5)}^{4}}}+%
\mathcal{C}_{2} e^{-\frac{\bar{\rho}_{0}}{6M_{(5)}^{4}}t},~~~(\omega _{b}=-1)
\label{eq:inflatingsol} \\
a_{0}(t) &=&2^{-\frac{1}{3(1+\omega _{b})}}\left[ const\pm \frac{(1+\omega
_{b})\bar{\rho}_{0}}{M_{(5)}^{4}}t\right] ^{\frac{1}{{3(1+\omega _{b})}}%
},~~~(\omega _{b}\neq -1)  \nonumber \\
&\sim &t^{\frac{1}{3(1+\omega _{b})}},  \label{eq:noninfl}
\end{eqnarray}%
where $\mathcal{C}_{1}$ and $\mathcal{C}_{2}$ are integration constants.

Eq. (\ref{eq:inflatingsol}) describes an inflationary solution for the minus
(plus) sign since $\bar{\rho}_{0}<0$ ($\bar{\rho}_{0}>0$) for the RS1 (RS2)
model. Although a detailed analysis of these solutions leads to a
non-conventional cosmology \cite{bdl}, the usual FRW universe can be
reproduced by various methods \cite{cgkt,cgrt,cgs,hbkim,bdel}. The crucial
difference between this and the previous brane-world model is that the
condensing ghost at late times cancels out the cosmological constant on the
brane, which can leads to the usual brane-world cosmology with a vanishing
cosmological constant.

Now consider the contribution of $%
\mathcal{P}(X)$ terms. Since this arbitrary kinetic function has a scale of $%
M_{(5)}^{-4}$ while the scale of the brane tension (pressure) terms is $%
M_{(5)}^{-8}$, we shall neglect the latter contribution. Furthermore, for a
slowly varying extra dimension, we can neglect terms proportional to $\ddot{r%
}_{0}$ in eqs. (\ref{eq:radii}) and (\ref{eq:combrel}). The equations of
motion then simplify to
\begin{eqnarray}
&&H_{0}^{2}+\frac{\alpha }{a_{0}^{2}}=\frac{4}{9}X\partial _{X}\mathcal{P}%
(X)-\frac{1}{9}\left( \mathcal{P}(X)-6k^{2}\right) ,  \label{eq:friedrad} \\
&&\frac{\ddot{a}_{0}}{a_{0}}=-\frac{4}{9}X\partial _{X}\mathcal{P}(X)-\frac{2%
}{9}\left( \mathcal{P}(X)-6k^{2}\right) ,  \label{eq:accelrad} \\
&&\frac{2}{3}(X\partial _{X}\mathcal{P}(X)-\mathcal{P}(X)+6k^{2})=3H_{0}%
\frac{\dot{r}_{0}}{r_{c}}.  \label{eq:radrel}
\end{eqnarray}%
The typical shape of the ghost condensate is shown in Fig. \ref{fig:1}.
We shall take $\mathcal{P}(X)$ to have the form%
\begin{equation}
\mathcal{P}(X)=\frac{1}{2}(X-c_{\ast }^{2})^{2}+6k^{2},
\label{eq:nearvacghost}
\end{equation}%
which should be approximately true near \ $X=c_{\ast }^{2}$ (i.e. in the
vicinity of the vacuum). With eq. (\ref{eq:nearvacghost}), the equations of
motion near the vacuum of the ghost condensate are
\begin{eqnarray}
&&H_{0}^{2}+\frac{\alpha }{a_{0}^{2}}=\frac{1}{18}(X-c_{\ast
}^{2})(7X+c_{\ast }^{2})\equiv \frac{8\pi }{3}\rho _{gc},  \label{eq:nvfried}
\\
&&\frac{\ddot{a}_{0}}{a_{0}}=-\frac{1}{9}(X-c_{\ast }^{2})(5X-c_{\ast
}^{2})\equiv -\frac{4\pi }{3}(\rho _{gc}+3p_{gc}),  \label{eq:nvaccel} \\
&&H_{0}\frac{\dot{r}_{0}}{r_{c}}=\frac{1}{9}(X-c_{\ast }^{2})(X+c_{\ast
}^{2}),  \label{eq:nvrad}
\end{eqnarray}%
where $\rho _{gs}$ and $p_{gs}$ represent respectively the energy density
and the pressure generated by the ghost condensate. The first two equations
describe the Friedmann and the acceleration equations while the third one is
the evolution equation of the extra dimension. Note that the first two
equations are similar to those shown in ref. \cite{kn} apart from some
factors. The fact that eqs. (\ref{eq:nvfried}) and (\ref{eq:nvaccel}) should
satisfy the conservation equation, $\dot{\rho}_{gc}=-3H_{0}(\rho
_{gc}+p_{gc})$, produces a differential equation for $X(t)$%
\begin{equation}
\dot{X}={\pm }\frac{(X-c_{\ast }^{2})(17X-c_{\ast }^{2})}{7X-3c_{\ast }^{2}}%
\sqrt{\frac{(X-c_{\ast }^{2})(7X+c_{\ast }^{2})}{18}},  \label{eq:xeqns}
\end{equation}%
and this equation is similar to that in ref. \cite{kn}.

For small perturbations,
the radion should be stable when $X$ goes to the condensing vacuum, which implies that
$r_{0}<<r_{c}$ should be valid for the solution of $r_{0}(t)$. Eqs. (\ref{eq:nvrad})
and (\ref{eq:xeqns}) becomes
\begin{equation}
  \label{eq:radevolvx}
  \frac{dr_{0}}{dX} = - \frac{2r_{c}(7X-3c_{\ast}^2)(X+c_{\ast}^2)}{(17X-c_{\ast}^2)(X-c_{\ast}^2)(7X+c_{\ast}^2)},
\end{equation}
which easily solves to
\begin{equation}
  \label{eq:radsolx}
  r_{0}(X) = \bar{r}_{0} - r_c\left(\frac{33}{136} \ln(17X-c_{\ast}^2) +
  \frac{1}{8}\ln(X-c_{\ast}^2) - \frac{1}{4}\ln(7X+c_{\ast}^2)\right),
\end{equation}
where $\bar{r}_{0}$ is an integration constant. The solution
(\ref{eq:radsolx}) shows that the radius of the extra dimension is destabilized
when $X$ goes to the stable vacuum, $X=c_{\ast}^2$, violating our assumption of
small radion perturbations.

As a consequence, Minkowski spacetime
at late times is not compatible with a stable radion field. To
resolve this, we shall in the next section turn on the bulk matter
along the $y$-direction, $p_{5}$, and show that this stabilizes the
radion field and alters the evolution of the scale factor and ghost field.

\section{Stable Radion and Cosmic Acceleration}
\label{sec:V}

In this section, we consider the full-set of equations
of motion with bulk matter along the $y$-direction, $p_{5}$. We show that this
can stabilize the radius of extra dimension and lead to the
exponentially inflating scale factor and an accelerating expansion
at late times. Introducing this sort of bulk matter has already shown
in refs. \cite{hbkim,kkop}, which turns out to be obviously responsible
for the radion stabilization.

\subsection{Late-Time Behaviors by the Bulk Matter}
We have three equations of motion for the slightly perturbed radius of
extra dimension with the bulk matter, $p_{5}$,
\begin{eqnarray}
  && H_{0}^2 + \frac{\alpha}{a_{0}^2} = - H_{0}\frac{\dot{r}_{0}}{r_{c}} +
  2 k^2 + \frac{1}{3}(2X\partial_{X}{\mathcal P}(X) - {\mathcal
  P}(X)),\label{eq:1.1}\\
  && H_{0}^2 + \frac{\alpha}{a_{0}^2}
  +2\left(\frac{\ddot{a}_{0}}{a_{0}} + H_{0}
  \frac{\dot{r}_{0}}{r_{c}}\right) = 6k^2 - {\mathcal P}(X),\label{eq:1.2}\\
  &&H_{0}^2 + \frac{\alpha}{a_{0}^2} + \frac{\ddot{a}_{0}}{a_{0}} =
  2k^2 - \frac{1}{3}{\mathcal P}(X) - \frac{1}{3}p_{5},\label{eq:1.3}
\end{eqnarray}
which are reduced to alternative form of equations by
\begin{eqnarray}
  && H_{0}^2 + \frac{\alpha}{a_{0}^2} = \frac{4}{9}X\partial_{X}{\mathcal P}(X)
  - \frac{1}{9}{\mathcal P}(X) + \frac{2}{3}k^2 - \frac{2}{9}
  p_{5},\label{eq:2.1}\\
  && \frac{\ddot{a}_{0}}{a_{0}} = - \frac{4}{9}X\partial_{X}{\mathcal P}(X) -
  \frac{2}{9}{\mathcal P}(X) + \frac{4}{3}k^2 -
  \frac{1}{9}p_{5},\label{eq:2.2}\\
  && H_{0}\frac{\dot{r}_{0}}{r_{c}} = \frac{2}{9} (X\partial_{X}{\mathcal P}(X) -
  {\mathcal P}(X) + 6k^2 + p_{5}).\label{eq:2.3}
\end{eqnarray}

On the other hand, the addition of $p_{5}$ should satisfies a
condition for stabilization \cite{kkop} given by $T^{\mu}_{~\mu} - 2 T^{y}_{~y} = 0$,
which leads to a general relation,
\begin{equation}
  \label{eq:genrelstable}
  {\mathcal P}(X) - X\partial_{X}{\mathcal P}(X) = 6k^2 + p_{5}.
\end{equation}
Assuming that $\partial_{X}{\mathcal P} \rightarrow 0$ and ${\mathcal P}
\rightarrow - D$ as $t\rightarrow\infty$, where $D$ is a constant from
the condensing ghost at $X=c_{\ast}^2$,
the set of equations becomes
\begin{eqnarray}
  && H_{0}^2 + \frac{\alpha}{a_{0}^2} = \frac{2}{3}k_{0}^2 -
  \frac{2}{9}p_{5},\label{eq:3.1}\\
  &&\frac{\ddot{a}_{0}}{a_{0}} = \frac{4}{3}k_{0}^2 -
  \frac{1}{9}p_{5},\label{eq:3.2}\\
  && H_{0}\frac{\dot{r}_{0}}{r_{c}} = \frac{2}{9} (6k_{0}^2 + p_{5}).\label{eq:3.3}
\end{eqnarray}
The stabilization condition (\ref{eq:genrelstable})
becomes $p_{5}+6k_{0}^2=0$, which leads to a simple result of the
stable radion, $r_{0} = const.$, from eq. (\ref{eq:3.3}). And we have
an inflationary solution of the scale factor for a flat ($\alpha=0$)
and the expanding universe ($k_{0}>0$),
\begin{equation}
  \label{eq:scalelatetimes}
  a_{0}(t) = \bar{a}_{0} e^{\sqrt{2}k_{0} t},
\end{equation}
where $\bar{a}_{0}$ is an integration constant. Note that $6k^2 =
\Lambda_{(5)}/M_{(5)}^4$ is negative (anti-de Sitter(AdS)), but the minimum value of
ghost condensate ($D>0$) renders the effective cosmological constant
($k_{0}^2 = D+6k^2$)
positive (dS phase), which ultimately can describe an accelerating
universe at late times. In other words, an initial AdS
spacetime becomes a dS spacetime at late times through the ghost
condensation mechanism. We will show
in the next section that ghost evolution also gives rise to this effect.
As seen before, the bulk field, $p_{5}$, stabilizes the radion at late
times and alters the evolution of the scale factor to be inflationary.

\subsection{Ghost Evolution and Accelerating Expansion of the Universe}

For the near vacuum of the condensing ghosts, ${\mathcal P}(X)=(X-c_{\ast}^2)^2/2-D$
, eqs. (\ref{eq:2.1}), (\ref{eq:2.2}), and (\ref{eq:2.3}) can be written
by
\begin{eqnarray}
  &&H_{0}^2 + \frac{\alpha}{a_{0}^2} =
  \frac{1}{18}(X-c_{\ast}^2)(7X+c_{\ast}^2) - \frac{2}{9}p_{5} +
  \frac{2}{3}k_{0}^2,\label{eq:12.1}\\
  &&\frac{\ddot{a}_{0}}{a_{0}} =
  -\frac{1}{9}(X-c_{\ast}^2)(5X-c_{\ast}^2)-\frac{1}{9}p_{5}+\frac{4}{3}k_{0}^2,
  \label{eq:12.2}\\
  &&H_{0}\frac{\dot{r}_{0}}{r_{c}} =
  \frac{1}{9}(X-c_{\ast}^2)(X+c_{\ast}^2) + \frac{2}{9}p_{5} + \frac{4}{3}k_{0}^2,\label{eq:12.3}
\end{eqnarray}
and eq. (\ref{eq:genrelstable}) can be evaluated by
\begin{equation}
\label{eq:stabrels}
  p_{5} = - 6k_{0}^2 - \frac{1}{2}(X-c_{\ast}^2)(X+c_{\ast}^2),
\end{equation}
which automatically leads to $r_{0}=const.$ from eq. (\ref{eq:12.3}).
It is easy to show that this result coincides with the one from the
energy-momentum conservation, $\nabla_{M}T^{MN}=
\nabla_{M}(T^{MN}_{\mathcal P} + T^{MN}_{\mathrm{bulk}}+T^{MN}_{\Lambda_{(5)}})=0$. The field equation for $X$ is equivalent to
$\nabla_{M}T^{MN}_{{\mathcal P}}=0$ and $\nabla_{M}(T^{Mt}_{\mathrm{bulk}}+T^{Mt}_{\Lambda_{(5)}})=0$ yields
\begin{equation}
  \label{eq:13}
  \frac{\dot{b}}{bn^2} p_{5} = 0,
\end{equation}
with the metric (\ref{eq:metric}). Here if we set $b=r_{c}+r_{0}(t)$ for the small perturbation,
we have $\dot{r_{0}} = 0$ for the non-vanishing $p_{5}$.
We therefore obtain two equations using eq. (\ref{eq:stabrels}),
\begin{eqnarray}
   &&H_{0}^2 + \frac{\alpha}{a_{0}^2} =
  \frac{1}{6}(X-c_{\ast}^2)(3X+c_{\ast}^2) +  2k_{0}^2,\label{eq:14.1}\\
  &&\frac{\ddot{a}_{0}}{a_{0}} =
  -\frac{1}{6}(X-c_{\ast}^2)(3X-c_{\ast}^2)+2k_{0}^2.
  \label{eq:14.2}
\end{eqnarray}
These equations are the same as those in ref. \cite{kn}, and the analysis is similar.
A plot of the deceleration parameter, $q_{0} = -\ddot{a}_{0}/a_{0}H_{0}^2$, and
the equation-of-state parameter, $\omega$ appear in fig.\ref{fig:2}.

The only drawback to this model is the rather ad-hoc appearance of $p_{5}$.
One might hope that it could arise from the ghost field in some manner. The only
possiblilty would appear to be giving the ghost field a $y$-dependence. However
it is clear that this proposal cannot succeed.
If we assume that $\phi=\phi(t,y)$, the stabilization condition becomes
\begin{eqnarray}
  &&(2M_{(5)})^{-1}\left(T^{\mu}_{~\mu} - 2T^{y}_{~y}\right) \nonumber \\
 &=& {\mathcal{P}}(X) - X\partial_{X}{\mathcal P}(X) -
 3\partial_{X}{\mathcal P}(X) \frac{(\phi')^2}{b^2} - 6k^2 = 0.  
 \label{eq:stcondy}
\end{eqnarray}
Comparing this with eq. (\ref{eq:genrelstable}), we have $p_{5} =
3\partial_{X}{\mathcal P}(X)(\phi')^2/{b^2}$. However,
this inevitably vanishes as $t$ goes to infinity since
$\partial_{X}{\mathcal P}(c_{\ast}^2) \rightarrow 0$ at late times, which
finally leads to a Minkowski vacuum after condensation, yielding again an
unstable radion field. Consequently the bulk field $p_{5}$ cannot
originate from the ghost. However we note that it can arise from
the back reaction of the dilaton coupling to the brane \cite{kkop}.

To summarize, introducing bulk matter stabilizes the
radius of the extra dimension, alters the evolution of the ghost
field and the scale factor, and prevents the cosmological constant from
vanishing, which renders a dS phase on the brane and finally leads to an
accelerating expansion at late times.

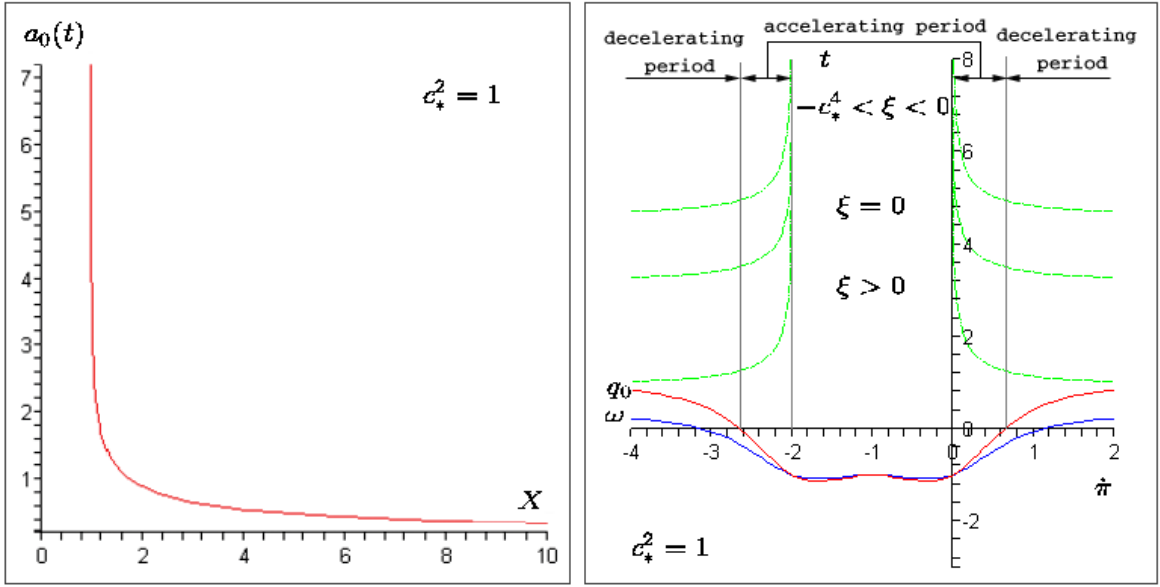
\fig{15.5cm}{8cm}{fig2}{\it The behavior
  of the scale factor, $a_{0}$, vs. $X$ (LHS) and the plot
  of the solution for $\dot{\pi}$ (green dash-dot line), the
  deceleration parameter (red solid line), and the equation-of-state
  parameter (blue solid line) for each $\xi=12k_{0}^2-c_{\ast}^4$,
  where $\xi>-c_{\ast}^4$ (RHS)}{fig:2}

\section{Higher-Derivative terms}

Higher derivative terms for the ghost, in which
${\mathcal P}(X)  = (X-c_{\ast}^2)^p+D$, where $p=1,2,3,\cdots$ can
be analyzed in a similar manner.  If $p_{5}=0$, we find that the situation can be
simply classified by two cases : one for $p=$even and another for $p=$odd.
For $p=$even, the general behavior is quite similar to the preceding case
for $p=2$ up to the factor $1/2$, which results from the fact that
there always exists a stable vacuum point at $X=c_{\ast}^2$.
However, the odd power cases do not include the stable
point at $X=c_{\ast}^2$, and so the situation is of no interest in connection
with the issue of the ghost condensation. Nevertheless, one might
consider two ways of circumventing the situation for $p=$odd. One is
to regard the ghost condensate vacuum $X=c_{\ast}^2$ as a metastable
state. Provided we can sit long enough at a point of inflection of
${\mathcal P}(X)$, these solutions could be viable for cosmological
evolution, with the true condensate vacuum occurring the global
minimum of ${\mathcal P}(X)$. Another way is to place the condensate
minimum on the brane, writing ${\mathcal P}(X) =
|X-c_{\ast}^2|^{2\gamma-1} +D$ where $\gamma=1,2,3,\cdots$. In this
case, the discontinuity at $X=c_{\ast}^2$ is identified with the
location of the brane at $y=0$. Since we also have a discontinuity at
$y=0$ we could construct a new model for $p=$odd power of ghosts that
possesses a stable minimum.

As before, this situation can be addressed by introducing
non-vanishing bulk matter, $p_{5}$, with no additional
modifications of the model. The bulk matter makes the radion
stable and alters the evolution of the ghost condensate, whose
equations reduce to
\begin{equation}
  \label{eq:diffarbp}
  \frac{dX}{dt} = \pm
  \left[\frac{2X(X-c_{\ast}^2)\sqrt{3(X-c_{\ast}^2)^{p-1}((2p-1)X+c_{\ast}^2)+18k_{0}^2}}{(2p-1)X-c_{\ast}^2}\right]
\end{equation}
or alternatively
\begin{equation}
  \label{eq:inaltformx}
  t-t_{0}=\int dX \frac{(2p-1)X-c_{\ast}^2}{2X(X-c_{\ast}^2)\sqrt{3(X-c_{\ast}^2)^{p-1}((2p-1)X+c_{\ast}^2)+18k_{0}^2}}
\end{equation}
which have not been previously analyzed. It is straightforward  to
show that the integrands of eq. (\ref{eq:inaltformx}) for each
$p\neq 1$ have qualitatively similar behavior. Consequently the
evolution of higher derivative ghosts with bulk matter is
qualitatively equivalent to the $p=2$ case for both $p$ even and
odd.

We can obtain some analytic information about the evolution of the
condensate in the large $t$ limit. Since $X$ goes to $c_{\ast}^2$
we can employ the ansatz $X=c_{\ast}^2 + f(t)e^{-\mu t}$, where
$\mu$ is a positive constant. This induces a
differential equation for $f(t)$ which is
\begin{equation}
  \label{eq:fdeq}
  \frac{df}{dt} = \left(\mu f\pm\frac{2\sqrt{3}f(c_{\ast}^2
  + fe^{-\mu t})\sqrt{(fe^{-\mu t})^{p-1}(2pc_{\ast}^2 + (2p-1)fe^{-\mu t})
  + 6k_{0}^2}}{2c_{\ast}^2(p-1)+(2p-1)fe^{-\mu t}}\right).
\end{equation}
For large $t$, exp$(-\mu t)$ goes to zero, which reduces the
differential equation to
\begin{equation}
  \label{eq:fdeq2}
  \frac{df}{dt} = \left[ \left(\mu \pm \frac{3\sqrt{2}k_{0}}{p-1}\right)f \mp
  \frac{3\sqrt{2}k_{0}f^2}{2(p-1)^2 c_{\ast}^2} e^{-\mu t}\right].
\end{equation}
To solve this, we note that $f$ should be a constant(i.e.
$df/dt=0$) at large $t$. This determines $\mu = \pm
3\sqrt{2}k_{0}/(p-1)$, where we take the plus sign to ensure a
decaying solution for $X$.

Now we consider the next leading term of eq.
(\ref{eq:fdeq2}). Plugging the value of $\mu$ into eq.
(\ref{eq:fdeq2}) leads to the solution,
\begin{equation}
  \label{eq:fsols}
  f(t) = \left(A + \frac{e^{-\frac{3\sqrt{2}k_{0}}{p-1}t}}{2c_{\ast}^2(p-1)}\right)^{-1}
\end{equation}
where $A>0$ is an integration constant. Eq. (\ref{eq:fsols}) shows
that the function $f(t)$ becomes a constant as $t$ goes to
infinity and the solution of $X$ is
\begin{equation}
  \label{eq:xsolus}
  X = c_{\ast}^2 + {e^{-\frac{3\sqrt{2}k_{0}}{p-1}t}}\left(A+\frac{e^{-\frac{3\sqrt{2}k_{0}}{p-1}t}}{2c_{\ast}^2(p-1)}\right)^{-1},
\end{equation}
illustrating that the solutions have qualitatively similar behavior. Since we are
working in the near-vacuum region, the effects of higher order
terms beyond the leading contribution at a given $p$ will be
negligible at late times.

\section{Conclusions}

\label{sec:VI}

We have studied the RS brane-world model with an arbitrary function $%
\mathcal{P}(X)$ of the kinetic term of a scalar field in the
five-dimensional AdS spacetimes. An interesting aspect of
our model is that the generic function $\mathcal{P}(X)$ is uniquely
determined in the brane-world with a static extra dimension. Once
time-evolution of the radius of the extra dimension is taken into account,
the solution gives rise to ghost condensation. This implies that the
excitation of the brane along the extra coordinate results in a ghost
condensate and the ghost field approaches the stable vacuum at late times.
Another intriguing feature of this model is that the minimum value
of the ghost condensate vacuum cancels out the cosmological
constant. This inevitably leads to the Minkowskian vacuum on the brane
after the ghost condensation at late times. In this background, a
radiation-dominated universe can be generated by the vacuum
fluctuations of the ghost condensate. As a result, vacuum fluctuations
of ghosts can generate radiating matter at late times by the ghost
condensation.

However, this scenario inevitably yields an unstable radion.
This instability can be circumvented by
introducing a bulk field along the $y$-direction, preserving the
consistency of the equations of motion and reducing the system
to that of the four-dimensional case \cite{kn}. This sort of bulk matter along
the extra dimension can arise from the back reaction of the dilaton
field \cite{kkop}, which is obviously responsible for the stabilized radion.
Since it alters the time-evolution of the scale factor and the ghost field,
the model ultimately leads to the inflationary expanding scale factor
and the acceleratingly expanding epoch at late
times. In addition, the condensing ghost behaves like a cosmological
constant since it approaches to $\omega = -1$ as times goes on.
One of the interesting features in this model is that the
spacetimes with an initially negative curvature (AdS) transfers to the
spacetimes with a positive curvature (dS) as the condensing ghost
approaches to the stable vacuum, which results from the fact that the
vacuum of the ghost condensation ($D$) forbids the effective cosmological
constant ($k_{0}^2$) to be negative or vanish.

For $p=2$ an analysis of our model has already been carried
out in ref. \cite{kn}. However there are some distinctions between
the two models. The main point of our model was to investigate
if/how ghost condensation can appear from a wiggling brane along
the extra dimension. One of the interesting features of our model
is that the zero point energy ($D$-term) of ghost condensation
inevitably must appear in order to preserve consistent equations
of motion, which are ultimately responsible for the dS expansion
at late times. The situation for larger $p$ has not been
previously analyzed; we find that the behaviour of the ghost
condensate is qualitatively similar to the $p=2$ case.

As a consequence, ghost condensation in the RS brane-world model
with the stabilized radius of the extra dimension provides
a possible explanation of an accelerating universe at late
times. In addition, the model might offer a possible
account of an early universe inflationary cosmology,
which would be worthwhile to explore in the future.

\newpage
\vspace{0.9cm}
\textbf{Acknowledgment}

We would like to thank P. Langfelder for helpful comments. JJO wishes to
thank H. S. Yang, I.-Y. Cho, W.-I. Park, and E. J. Son for invaluable
discussions and P. S. Apostolopoulos for useful comments and
discussions. JJO was supported in part by the Korea Research Foundation
Grant funded by Korea Government (MOEHRD, Basic Research Promotion Fund
No. M01-2004-000-20066-0). This work was supported in part by the Natural
Sciences \& Engineering Research Council of Canada.


\end{document}